# "Clumps" in the Resistive-Drift-Wave turbulence

S. I. Krasheninnikov and R. D. Smirnov

*University of California San Diego, La Jolla, CA 92093, USA*

**Abstract**

The results of numerical simulations of the Hasegawa-Wakatani equation demonstrate that, similarly to decaying turbulence in 2D fluids, at a small electron adiabaticity parameter, the resistive-drift-wave (RDW) turbulence is dominated by the vortices. Occasionally, vortices with different signs become coupled and propagate ballistically as a dipole over a large distance, entraining plasma density. Such ballistic motion of the vortex-density "clumps" in radial direction provides a non-local feature of plasma transport associated with the RDW turbulence. Large magnitude of plasma parameters perturbations associated with clumps can initiate other plasma instabilities and nonlinear phenomena.

## I. Introduction

The mesoscale structures, such as "avalanches" [1-5] and "blobs" [6-10], exhibiting the ballistic radial motion, play an important role in the transport of particles and energy of the magnetized fusion plasma. They not only contribute to the magnitude of anomalous plasma fluxes, but also provide the non-local features to it. Even though the importance of these mesoscale structures in the magnetized plasma transport was recognized a long time ago, the physics of such objects has many open issues, and the extensive studies of these phenomena, both theoretical and experimental, continue (e.g. see [11-13] and the references herein).

The avalanche is usually portrayed as a radial "front" of perturbed plasma parameters and plasma (particle/energy) flux propagating "ballistically" in radial direction. As a "proof" of such a phenomenon, following from the simulations, the researchers often present the poloidally averaged flux, j(t,x), as a function of time t and the radial coordinate x. For the case of the radial ballistic propagation of the incremental flux, δj, the function δj(t,x) produces a straight "stripe" on (t, x) plane schematically depicted in Fig. 1 (see also Fig. 2 from [4]).

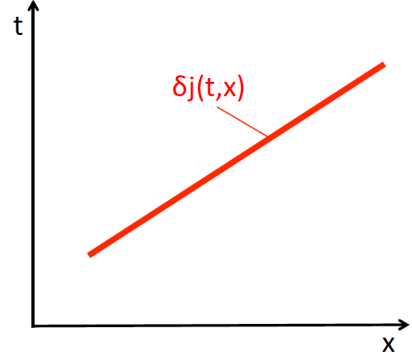

Fig. 1. The function of the incremental flux δj(t,x) driven by the avalanche.

Interestingly, the results of our study of the Resistive Drift Wave (RDW) turbulence also find the ballistically advected nonlinear structures, which, however, have limited extent in both radial and poloidal directions. Untangling the physics of these structures is the subject of this work.

The paper is organized as follows: In Section II we present the dimensionless form of the modified Hasegawa-Wakatani (mHW) equations [14, 15] and describe the technique used for numerical simulations; in Section III we discuss the results of our numerical simulations, introduce the concept of "clumps", the dynamics of which exhibits the ballistic features and give a crude analytic estimates of their impact on the overall anomalous plasma transport driven by the RDW turbulence; in Section IV we summarize our findings and discuss their implications.

## II. Equations and numerical approach

The mHW equations adopt the cold ion approximation and assume fixed electron temperature, $T_e$, whereas the plasma density, $N(\vec{r},t)$, can have a small departures from the equilibrium stationary value, $\bar{N}(x)$, which gives $N(\vec{r},t) = \bar{N}(x) + \hat{n}(x,y,t)$ where $|\hat{n}|/\bar{N} \ll 1$. In the dimensionless form the mHW equations can be written as follows

$$\frac{\partial}{\partial t}\nabla_\perp^2 \phi + \{\phi, \nabla_\perp^2 \phi\} = \alpha(\tilde{\phi} - \tilde{n}) + \hat{D}_\phi(\nabla_\perp^2 \phi), \quad (1)$$

$$\frac{\partial n}{\partial t} + \{\phi, n\} = \alpha(\tilde{\phi} - \tilde{n}) - \frac{\partial \phi}{\partial y} + \hat{D}_n(n), \quad (2)$$

where $\{a,b\} \equiv \partial_x a \partial_y b - \partial_y a \partial_x b$ is the Poisson bracket; "x" and "y" are the "radial" and "poloidal" coordinates; $\tilde{a} = a - \langle a \rangle_y$ and $\langle a \rangle_y$ means the averaging over coordinate y. The following normalizations were implemented in Eq. (1, 2): $\vec{r}_\perp / \rho_s \to \vec{r}_\perp$, $\rho_s^2 = T_e / (M\Omega_{Bi}^2)$, $\Omega_{Bi} = eB/Mc$, e is the elementary charge, B is the magnetic field strength ($B > 0$), M is the ion mass and c is the light speed; $\kappa \Omega_{Bi} t \to t$, $\kappa = -\rho_s d\ell n(\bar{N}_e(x))/dx = \text{const.} > 0$; $n = (\hat{n}/\bar{N})/\kappa$; $\phi = (e\varphi/T_e)/\kappa$ where $\varphi$ is the electrostatic potential; and the electron adiabaticity parameter

$$\alpha = (T_e/m)k_\parallel^2 / (\nu_{ei}\Omega_{Bi}\kappa), \qquad (3)$$

where $k_\parallel = \text{const.}$ is the component of the wave number parallel to the magnetic field lines, m is the electron mass, and $\nu_{ei}$ is the electron-ion Coulomb collision frequency.

To describe the small-scale "dumping" effects, which are beyond our approximation, and to improve the numerical stability of our simulations, we add the dissipative terms $\hat{D}_a(a)$ (e.g. see Ref. 16) to Eqs. (1, 2). The numerical solutions of Eq. (1, 2) were performed with a pseudo-spectral code Dedalus [17] using Fourier basis functions with double periodic boundary conditions on an (x, y) square box with the size L (in dimensionless units). The simulation domain contained 512×512 grid points with the smallest resolved wavenumber $\delta k = \delta k_x = \delta k_y = 0.1$ and the box size $L = 2\pi/\delta k$. The 3/2 de-aliasing rule was used [18]. The simulations continued for $10^3$ dimentionless time units, where all the results presented below correspond to the fully developed non-linear turbulence stage.

## III. The results of numerical simulations and analytic estimates

It is well known that the statistically averaged normalized plasma flux described by the mHW equations, $\Gamma = \langle \tilde{n} \partial_y \tilde{\phi} \rangle$, exhibits a strong reduction when the adiabaticity parameter goes from $\alpha < 1$ to $\alpha > 1$ [15]. The reason for such a reduction is the formation of the patterns of stationary zonal flows distributed in radial directions, which trap the drift wave turbulence in the radially localized effective potential wells, preventing their interactions and, therefore, reducing anomalous transport (e.g. see [19, 20]).

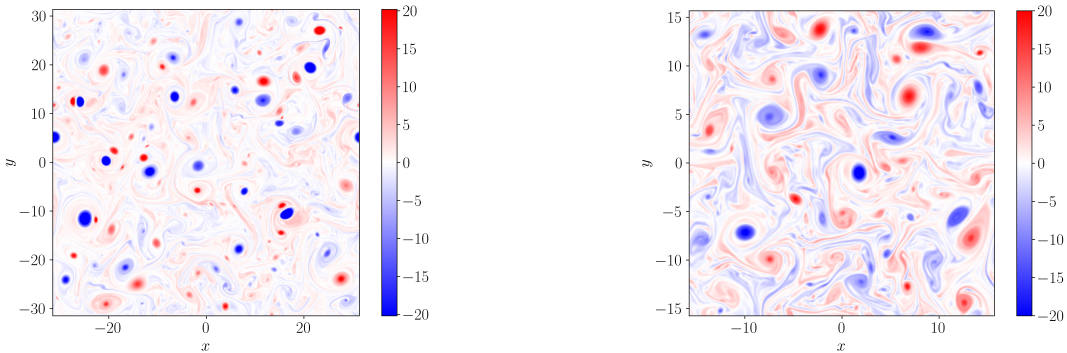

Fig. 2. The snapshot of vorticity, $\zeta \equiv \nabla^2 \phi$, found from numerical simulations for $\alpha = 10^{-2}$ (left) and $\alpha = 10^{-1}$ (right). Notice different scales of the x, y coordinates.

However, for $\alpha < 1$ no strong zonal flow emerges and the RDW turbulence is dominated by the very long-lived vortices characterized as localized structures of vorticity, $\zeta \equiv \nabla^2 \phi$, shown in Fig. 2, which found from our simulations of Eq. (1, 2) for $\alpha = 10^{-2}$ and $\alpha = 10^{-1}$ (see also Ref. 21). We notice that for small $\alpha$ the vortex-dominated RDW plasma turbulence becomes very similar to the decaying turbulence in 2D fluid, which is also dominated by the long-lived vortices (e.g. see Ref. 22, 23 and the references therein).

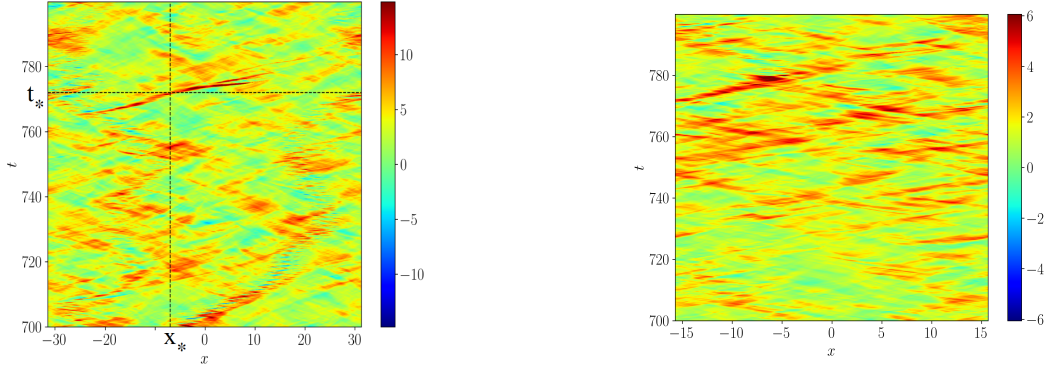

Fig. 3. The function $j(t,x) = \langle \tilde{n} \partial_y \tilde{\phi} \rangle_y$ found from numerical simulations of Eq. (1, 2) for $\alpha = 10^{-2}$ (left) and $\alpha = 10^{-1}$ (right)

Even though Fig. 2 does not reveal any visible front-like characteristics, we find that the poloidally averaged plasma flux $\hat{j}(t,x) \equiv \langle \tilde{n} \partial_y \tilde{\phi} \rangle_y$ driven by the RDW turbulence and shown in Fig. 3 exhibits the ballistic-like features similar to that of Fig. 1, indicating the advection-like contribution to plasma transport.

More close examination of the results of our RDW turbulence simulations reveals that the origin of the "stripes" on the function $\hat{j}(t,x)$, is "pairing" of the vortices having different signs but similar magnitude. As a result, they form poloidally oriented "dipoles", which propagate ballistically over a large distance and "drag" with them the whole plasma density. Such vortex-density "clumps" moving down the hill of $\bar{N}(x)$ build up the density bump, advected with the pair, whereas the pairs moving up the hill build and advect the density void (see the "stripes" in Fig. 3, which "move" to the left along x-axis in time). Eventually, the clumps are disintegrated due to the interactions with other vortices and small-scale turbulence.

Such clump can be clearly seen on the snap-shots of the plasma flux $j(t,x,y) \equiv n \partial_y \tilde{\phi}$ (at time $t = t_*$ corresponding to the horizontal dashed line in Fig. 3 (left)), as a vortex pair, and a density bump, which are encircled in black in Fig. 4.

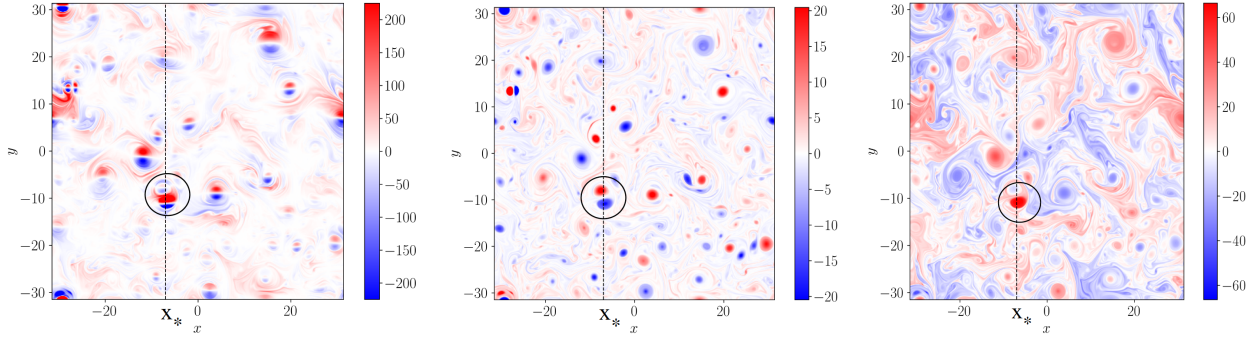

Fig. 4. The snap-shots of the flux $j(t,x,y) \equiv n\partial_y\tilde{\phi}$ (left), vorticity $\zeta$ (center), and density $n$ (right) found from numerical simulations of Eq. (1, 2) for $\alpha = 10^{-2}$ and time $t = t_*$ corresponding to the horizontal dashed line in Fig. 3 (left).

To demonstrate the clump's advection, in Fig. 5 one can see the snapshots of $j(t,x,y)$ in the three consecutive time moments where the same clamp is encircled in black.

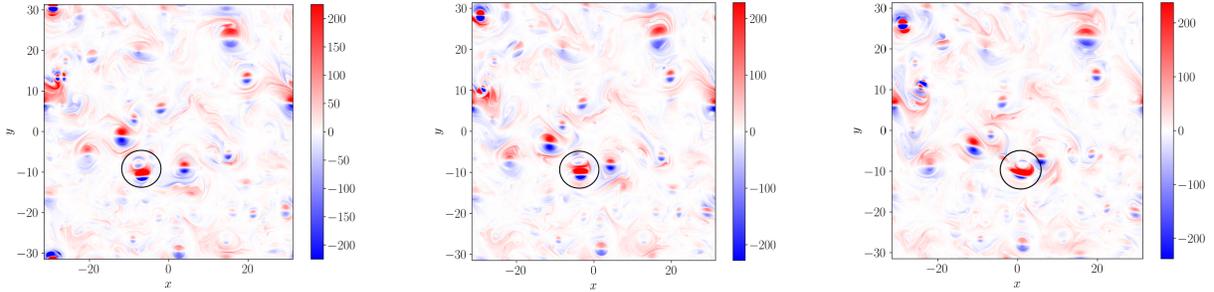

Fig. 5. The snapshots of $j(t,x,y)$ in three consecutive times where the same clamp, which advects in a radial direction, is encircled in black ($\alpha = 10^{-2}$).

To demonstrate that the clump's contribution to $\hat{j}(t,x)$ at time $t_*$ and x-coordinate $x_*$, corresponding to the dashed lines in Fig. 3(left) and the dashed line going through the clump in Fig. 4 (left), is dominant, in Fig. 6 we are showing the function $J(y) = L^{-1}\int_{-L/2}^{y} j(t_*,x_*,y')dy'$.

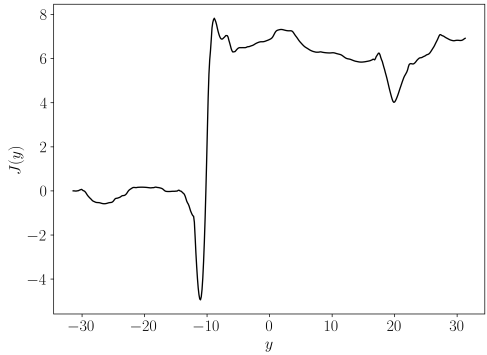

Fig. 6. The function $J(y)$, found from numerical simulation for $\alpha = 10^{-2}$.

As we see from the jump of the magnitude of $J(y)$ when the coordinate y goes though the clump, virtually all contribution to $J(y = L/2) = \hat{j}(t_*,x_*)$ shown in Fig. 3 (left), comes from the clump.

To evaluate the contribution of clumps to the statistically averaged plasma flux $\Gamma$, driven by the RDW turbulence, in Fig. 7 we present the time-averaged probability distribution function, $P_j(j)$, of the flux $j(t,x,y)$ found from numerical simulations for $\alpha = 10^{-2}$.

The "wings" of $P_j(j)$ correspond to the clumps. Since clumps, as it can be seen in Fig. 4, include parts with both negative and positive values of the flux $j(t,x,y)$ we account for the clump contribution considering the function $\hat{\Gamma}(j_0) = \int_{|j| \geq j_0} P_j(j) j dj$, shown in Fig. 8. We notice that $\hat{\Gamma}(j_0 = 0) \equiv \Gamma$. As we see from Fig. 8, the contribution of clumps, corresponding to the tail of the function $\hat{\Gamma}(j_0)$, to the overall plasma flux although sizable, but is within ~10%.

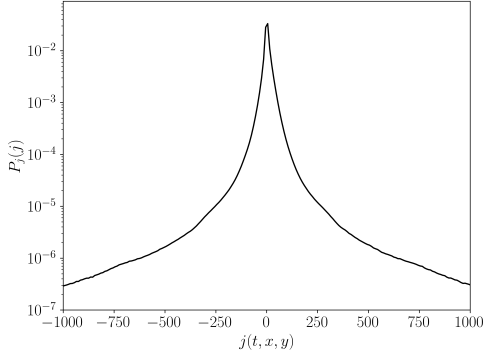
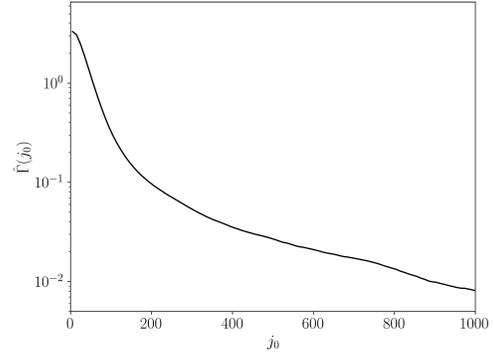

Fig. 7. The time averaged probability distribution function of the flux $j(t,x,y)$ found from numerical simulations for $\alpha = 10^{-2}$.

Fig. 8. The function $\hat{\Gamma}(j_0)$ for $\alpha = 10^{-2}$.

To make some qualitative estimates of contribution of clumps to the RDW driven anomalous plasma transport we notice that from Eq. (1, 2) one finds the following estimates (in our dimensionless units) of the characteristic frequency, $\omega$, the turbulence correlation time, $\tau_{cor}$, the wave number, $K_\perp$, the magnitudes of the fluctuating electrostatic potential, $\tilde{\phi}$, and density, $\tilde{n}$, for $\alpha < 1$ (see also Ref. 24 and the references therein):

$$\omega \sim \tau_{cor}^{-1} \sim K_\perp \sim \alpha^{1/3}, \quad \tilde{n} \sim \tilde{\phi} \sim \alpha^{-1/3}. \qquad (4)$$

From Eq. (4) we find the following estimate for the effective anomalous diffusion coefficient, $D_{RDW}$, and anomalous flux, $\Gamma$:

$$D_{RDW} \sim K_\perp^{-2} \tau_{cor}^{-1} \sim \alpha^{-1/3}, \quad \Gamma \sim \tilde{n} K_\perp \tilde{\phi} \sim \alpha^{-1/3}. \qquad (5)$$

We notice that the numerical simulations in Ref. 21 confirm, in a ballpark, this scaling of $\Gamma$.

From "dimensional" reasons we can argue that the "density" of the vortices on the (x, y) plane, $N_{vort}^{(2D)}$, can be evaluated as $N_{vort}^{(2D)} \sim K_\perp^2 \sim \alpha^{2/3}$. However, the vortex pairing only happens when the vortices approach each other on the distance $\sim \xi K_\perp^{-1}$, where $\xi < 1$ is some fitting parameter. Therefore, the density of clumps, $N_{clump}^{(2D)}$, on the (x, y) plane can be estimated as $N_{clump}^{(2D)} \sim \xi^2 K_\perp^2 \sim \xi^2 \alpha^{2/3}$, whereas the velocity of the clump advection, $V_{clump}$, as $V_{clump} \sim \xi^{-1} K_\perp \tilde{\phi} \sim \xi^{-1}$. As a result, the contribution of clumps to anomalous plasma flux, $\Gamma_{clump}$, can be estimated as

$$\Gamma_{clump} \sim N_{clump}^{(2D)} \tilde{n} V_{clump} \sim \xi \alpha^{1/3}. \tag{6}$$

Thus, for the relative contribution of the clumps to anomalous plasma transport we find

$$\Gamma_{clump} / \Gamma \sim \xi \alpha^{2/3} < 1. \tag{7}$$

Comparing the estimate (7) with the simulation data from Fig. 8, we find that transition to the clump induced tail of the function $\hat{\Gamma}(j_0)$ found for $\alpha = 10^{-2}$ is, in a ballpark, consistent with Eq. (7).

**IV. Conclusions**
From the modified Hasegawa-Wakatani equations, describing the RDW turbulence, we find that for a small electron adiabaticity parameter $\alpha$, pairing of the turbulence vortices produces rather long-living dipolar structures, which propagate ballistically over large distances and entrain large plasma density perturbations. Such "clumps" produced by the RDW turbulence and consisting of vortex pairs and density perturbations cause radial plasma convection similar to that of the curvature-driven blobs [6-10]. Interestingly, in the RDW clumps, radial advection of the vortex pair produces plasma density perturbation, whereas in curvature-driven blobs, density perturbations produce a vortex pair, which advects the whole structure.

Thus, clumps can explain experimental results on intermittent convective structures observed in linear devices [25, 26], where the curvature-driven mechanism is absent. For the cases where curvature effects are present, for example, in tokamaks and stellarators, the RDW turbulence-induced clumps also can be important for the under-threshold excitation of the ballooning filaments [27] and serve as the seeds for the curvature-driven blobs/filaments observed in H-mode plasmas [28, 29].

**Author Declarations**
**Conflicts of Interest**
The authors have no conflicts to disclose.

**Data Availability**
The data that support the findings of this study are available from the authors upon reasonable request.